\documentclass[11pt]{article}
\usepackage{blois,epsfig}

\def\be{\begin{equation}}
\def\ee{\end{equation}}
\def\bea{\begin{eqnarray}}
\def\eea{\end{eqnarray}}

\begin{document}
\vspace*{4cm}
\title{Heavy Flavour Spectroscopy}

\author{E.S. Swanson}

\address{Department of Physics and Astronomy, University of Pittsburgh, Pittsburgh PA 15260}

\maketitle\abstracts{
Recent issues in heavy flavour physics are reviewed. Anomalies in charmonium, $B$, and $\Upsilon(5S)$ decays and production are highlighted.  New results concerning exotic heavy quark states are also briefly reviewed.
}

\section{Introduction}

One often hears opposing views concerning hadrons. The first holds that hadrons are a reflection of ``irreducible complexity", with the implication that they are so difficult to understand that little can be gained by studying them. The second view holds that hadrons are ``simple", $SU(6)$ and perhaps some perturbative gluon exchange are enough  to understand them. The reality is more subtle and interesting than either of these simplistic positions.

At a basic level one hopes that hadronic spectroscopy will reveal insight into a highly nontrivial and nonperturbative field theory. One could argue, in fact, that any attempts at constructing BSM models are hopeless without a prior ability to understand QCD. At a more introspective level, hadron spectroscopy serves as an entree into the study of nonperturbative phenomena such as colour confinement, chiral symmetry breaking, topological excitations, and gluonic excitations. Looking further afield, it has possible BSM applications in CP or lepton flavour violation in $J/\psi$ decays, or Higgs effects in $\Upsilon$ decays. BSM physics has even been implicated in the $D_s$ decay constant \cite{Ds}.

\section{Perturbative QCD}

The main tools being applied to hadrons and their interactions are models; effective field theories such as chiral perturbation theory, soft collinear effective field theory (SCET), nonrelativistic QCD (NRQCD),  potential NRQCD (pNRQCD); and lattice gauge theory. These methods supersede older perturbative QCD computations of things such as quarkonium decay rates. However there is an impression that these older calculations have essentially solved many problems.  In fact the situation is much more nuanced, with many naive computations simply failing to explain the data. Since it is important to know where our models fail, I will describe several current issues here.

\subsubsection{The $\pi\rho$ Puzzle}

The $\pi-\rho$ puzzle is a longstanding issue in $J/\psi$ decays. The idea is that all decays of $J/\psi$ must proceed via $c\bar c$ annihilation, and therefore the wavefunction at the origin. Assuming that the same applies for  the $\psi'$ then permits the simple expectation \cite{ap}:

\be
\frac{Bf(\psi' \to h)}{Bf(J/\psi \to h)} = \frac{Bf(\psi' \to e^+e^-)}{Bf(J/\psi \to e^+e^-)} \approx 12.7\%
\ee

The experimental situation is illustrated in 
Fig. 1. 
As can be seen most points lie within a (sometimes large) standard deviation of 13\%, with the exception of the $\pi\rho$ final state, which is spectacularly far from expectations. The explanation for this remains elusive \cite{pirho}, with most explanations focussing on final state effects. But perhaps this is telling us that we do fully understand quark annihilation.

\begin{figure}[ht]
\begin{center}
\includegraphics[width=8.0cm,angle=0]{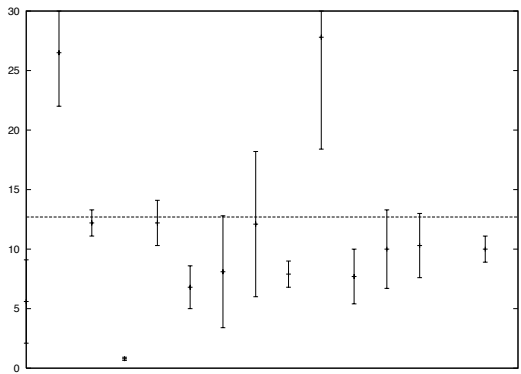}
\caption{The ratio $Bf(\psi' \to h)/Bf(J/\psi \to h)$ for the following final states (left to right): $K\bar K$, $p\bar p$, $\pi\rho$, $p\bar p \pi$, $4\pi$, $6\pi$, $3(\pi^-\pi^-)\pi^0$, $2(\pi^-\pi^-)\pi^0$, $2(\pi^-\pi^-\pi^0)$, $2(K^+K^-)$, $K^+K^-\pi^+\pi^-$, $K^+K^-\pi^+\pi^-\pi^0$, $K^+K^-4\pi$, $p\bar p \pi^+\pi^-$.}
\end{center}
\label{pirhoFig}
\end{figure}

\subsubsection{$e^+e^-$ Widths}

A similar story is played out in the $e^+e^-$ widths of charmonia. A simple and old model, due to van Royen and Weisskopf, is supposed to capture this physics (and is essentially unchanged in modern effective field theory approaches). The idea, again, is that transitions occur via the wavefunction at the origin, giving specific formulas:

\be
\Gamma(^3S_1 \to e^+e^-) = 16 \alpha_s^2 Q^2 \frac{|\psi(0)|^2}{M^2}
\ee
and

\be
\Gamma(^3D_1 \to e^+e^-) = 50 \alpha_s^2 Q^2 \frac{|\psi''(0)|^2}{M^2m_c^4}.
\ee
If this idea is right, quark model computations should be reasonably reliable. However comparison with experiment reveals a situation far from ideal (see Table \ref{vRWTab}). Although all the model predictions are in the right range, none of them are particularly successful. The $\psi(3770)$ rate prediction is especially poor. A possible explanation would be $S-D$ wave mixing, but detailed computations find it difficult to achieve mixing of the degree required \cite{BGS}. Again, something seems to be wrong our ideas of quark annihilation\footnote{There is a caveat here. Experience shows that $e^+e^-$ widths are very sensitive to the amplitude model used to fit the experimental data.}

\begin{table}[h]
\caption{Theoretical and Experimental $e^+e^-$ Charmonium Widths}
\vspace{0.4cm}
\begin{center}
\begin{tabular}{|llcc|}
\hline
state & model & theory (keV)  & expt (keV) \\
\hline
$J/\psi$     & $1^3S_1$ & 12 & 5.40(17) \\
$\psi'$      & $2^3S_1$ & 5   & 2.12(12) \\
$\psi(3770)$ & $1^3D_1$ & 0.06  & 0.26(4)\\
$\psi(4040)$ & $3^3S_1$ & 3.5 & 0.75(15) \\
$\psi(4159)$ & $2^3D_1$ & 0.1 & 0.77(23) \\
$\psi(4415)$ & $4^3S_1$ & 2.6 & 0.47(10) \\
\hline
\end{tabular}
\end{center}
\label{vRWTab}
\end{table}

\subsubsection{Radiative Charmonium Transitions}

As with the other topics covered so far, heavy quarkonium decay to photons is supposed to be governed by computable short distance physics. Employing this insight led to an old result for the ratio of $\chi_c$ decays \cite{Bar}:

\be
R = \frac{\Gamma(\chi_{c2} \to \gamma\gamma)}{\Gamma(\chi_{c0}\to \gamma\gamma)} = \frac{4}{15}(1 - 1.76\, \alpha_s) = 0.12 \ (\alpha_s = 0.32).
\ee
This prediction should be reasonably robust since unknown dynamics are eliminated in the ratio. However comparison with recent data from CLEO indicate something of a failure \cite{cleo}

\be 
R = \frac{0.66 \pm 0.07 \pm 0.04 \pm 0.05 \ {\rm keV}}{2.36 \pm 0.35 \pm 0.11 \pm 0.19 \ {\rm keV}} = 0.278 \pm 0.050. 
\ee
It is amusing, however, to note that the naive nonrelativistic quark model prediction for this ratio is simply $4/15 \approx 0.27$. Perhaps the perturbative asymptotic series is especially misleading in this case.

In a similar fashion, recent experiment reveals that \cite{pedlar}

\be
Bf(J/\psi \to \gamma\gamma\gamma) = (1.17 \pm 0.3 \pm 0.1)\cdot 10^{-5}
\ee
This result is in agreement with leading order pQCD, but the next to leading order result is negative.

Finally we draw attention to the following curious decay rate ratios \cite{pedlar2}:

\be
\frac{Bf(J/\psi \to \gamma\eta)}{Bf(J/\psi \to \gamma\eta')} = \frac{11.01 \pm 0.29 \pm 0.22}{52.4 \pm 1.2 \pm 1.1} = 0.21 \pm 0.04
\ee

\be
\frac{Bf(\psi(2S) \to \gamma\eta)}{Bf(\psi(2S) \to \gamma\eta')} = \frac{< 0.02}{1.19 \pm 0.08 \pm 0.03} < 0.018
\ee
What could be driving an order of magnitude  difference between these rates?

\subsubsection{Other Charmonium Transitions}

\begin{figure}[ht]
\includegraphics[width=8.0cm,angle=0]{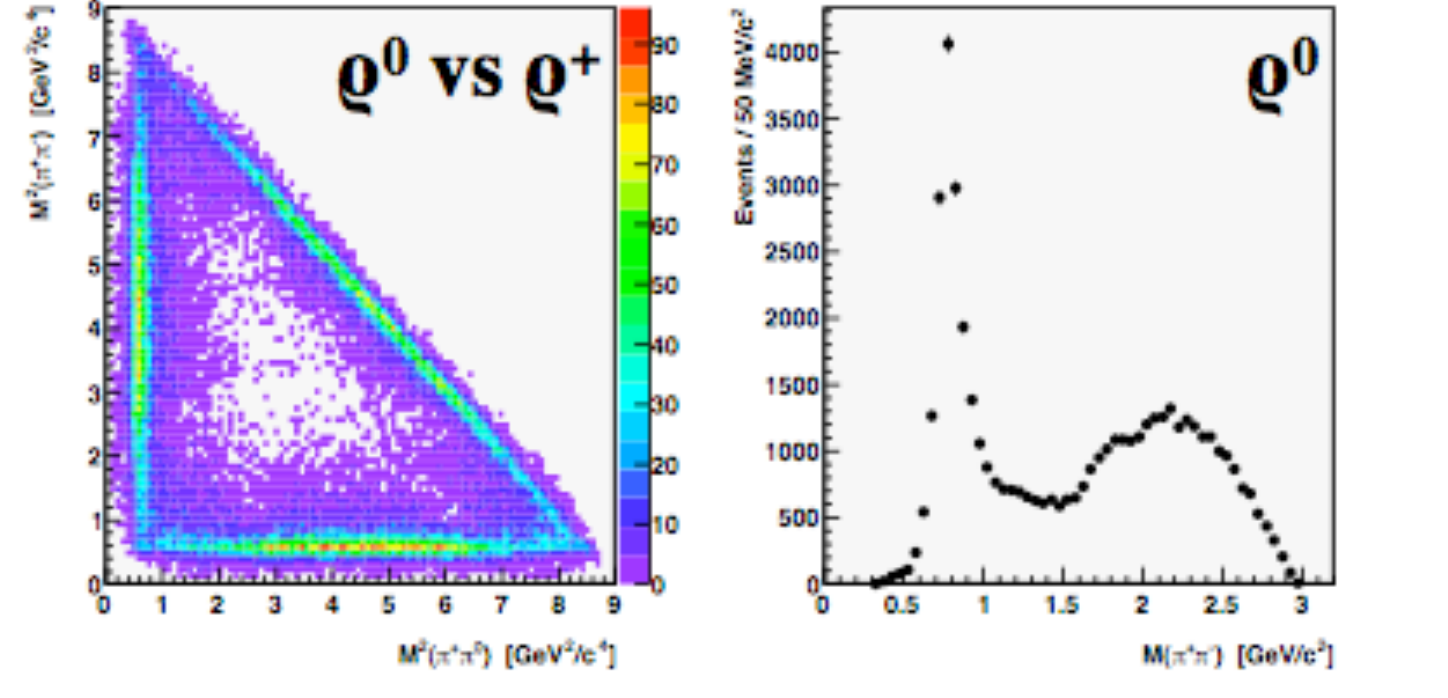}
\qquad
\includegraphics[width=8.0cm,angle=0]{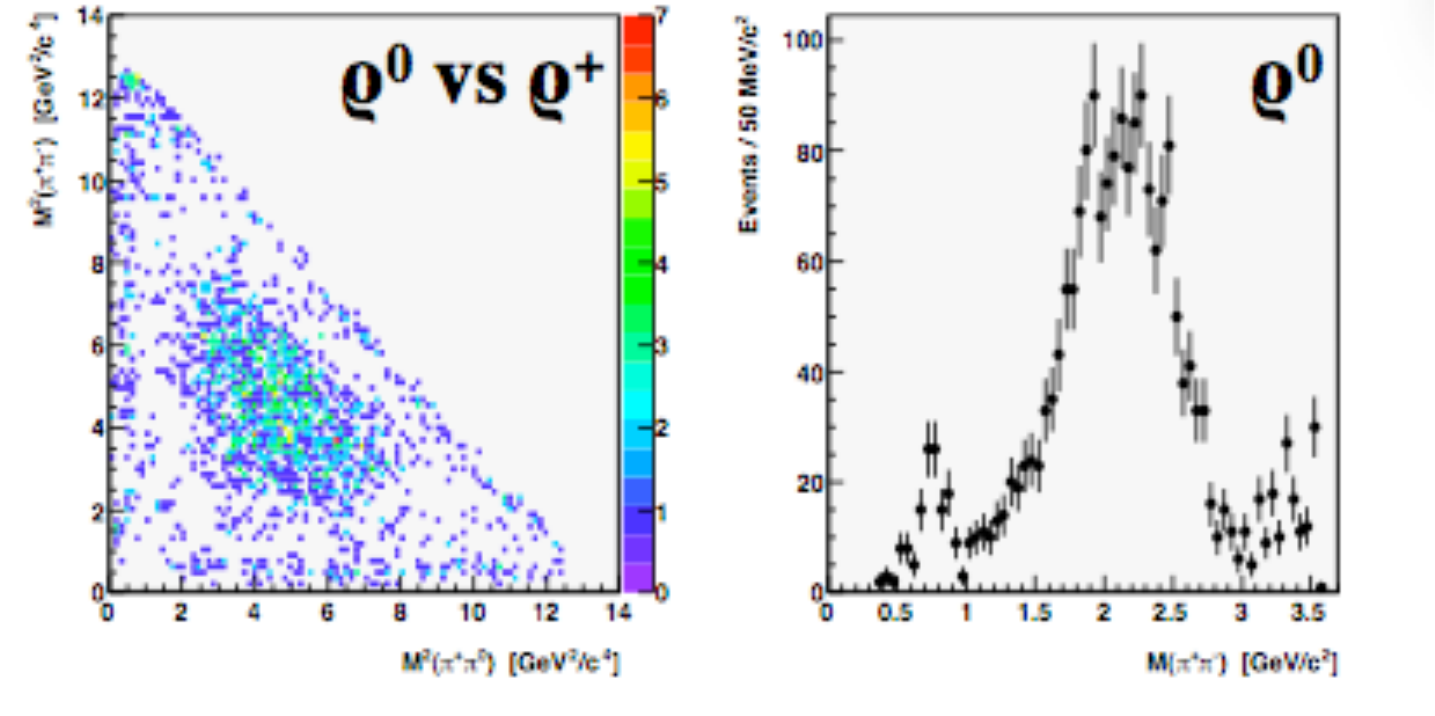}
\caption{$J/\psi \to \pi\pi\pi$ Dalitz Plot (left) and $\psi(2S) \to \pi\pi\pi$ (right).}
\label{psiDecayFig}
\end{figure}

Fig. \ref{psiDecayFig} displays Dalitz plots for the decays $J/\psi \to 3\pi$ and $\psi(2S)\to 3\pi$ \cite{psi3pi}. The difference between the plots is striking: a prominent $\rho$ is seen in the $J/\psi$ Dalitz plot while it is the $\rho'$ that dominates the $\psi(2S)$ decay. It seems unlikely that the extra energy available to the pions in $\psi(2S)$ decay could yield such a large difference, and the explanation must be sought in dynamics. In fact the data appears to telling us that the radial structure of the $\psi(2S)$ survives the annihilation into gluons to re-emerge in the final state. This is indeed a strange situation that deserves careful study.

The $\psi(2S)$ also figures in the following curious observation \cite{ryan}
\be
\frac{Bf(J/\psi \to \omega \eta)}{Bf(J/\psi \to \omega \eta')} = 9.56 \pm 0.16
\ee
yet
\be
\frac{Bf(\psi(2S) \to \omega \eta)}{Bf(\psi(2S) \to \omega \eta')} < 0.343.
\ee
Either the extra energy in the final state or the radial structure of the charmonium appears to have a dramatic effect on decay rates. 

Similar anomalies appear in $B$ decays. For example, one has \cite{blanc}

\be
\Gamma(B \to \eta K) \gg \Gamma(B\to \eta' K)
\ee
but
\be
\Gamma(B \to \eta K^*) \ll \Gamma(B\to \eta' K^*).
\ee
Could a simple spin flip in the kaon cause this behaviour?

Lastly, the production of charmonia recoiling off of $J\psi$ in $e^+e^-$ reactions remains problematic. Experiment and (older) theory for this process are given in Table \ref{recoilTab}. The theoretical computations were made with NRQCD (``BL") or variants. While higher order corrections tend to improve the agreement, they still do not satisfactorily explain the data. More importantly, this is a strong indication that NRQCD is not a good starting point for describing some processes.

\begin{table}[h]
\caption{Cross Sections (fb) for $e^+e^- \to J/\psi H$ at $\sqrt{s} = 10.6$ GeV.}
\vspace{0.4cm}
\begin{center}
\begin{tabular}{|lccc|}
\hline 
 $H$:  & $\eta_c$  & $\chi_{c0}$ & $\eta_c'$ \\
\hline
BaBar \cite{bernard} &  $17.6 \pm 2.8 \pm 2.1$ & $10.3 \pm 2.5 \pm 1.8$ & $16.4 \pm 3.7 \pm 3.0$ \\
Belle \cite{belleD} &  $25.6 \pm 2.8 \pm 3.4$ & $6.4 \pm 1.7 \pm 1.0$ & $16.5 \pm 3.0 \pm 2.4$ \\
\hline
BL \cite{BL} &  $2.31 \pm 1.09$ & $2.28 \pm 1.03$ & $0.96 \pm 0.45$ \\
LHC \cite{LHC} & 5.5                 & 6.9             & 3.7            \\
BC \cite{Bondar:2004sv} & $\sim 33$ &  &  \\
BLL \cite{BLL} & 26.7 & & 26.6 \\
\hline
\end{tabular}
\end{center}
\label{recoilTab}
\end{table}

\subsubsection{$\Upsilon(5S)$ Decays}

The $B$ factories have provided a wealth of information on electroweak transitions and $\Upsilon$ decays.  Recent data on $\Upsilon(5S)$ decays reveals several anomalies. For example the ratio of decay rates \cite{mizuk}:

\be
\frac{\Gamma(\Upsilon(5S) \to h_b(1P) \pi\pi)}{\Gamma(\Upsilon(5S) \to \Upsilon(2S)\pi\pi)} = 0.407 \pm  0.079 \pm 0.06
\ee

\be
\frac{\Gamma(\Upsilon(5S) \to h_b(2P) \pi\pi)}{\Gamma(\Upsilon(5S) \to \Upsilon(2S)\pi\pi)} = 0.78 \pm  0.09 \pm 0.15
\ee
indicate that transitions involving heavy quark spin flip (the $h_b$ is dominantly $S=0$) are not suppressed, in seeming violation of heavy quark effective field theory. An explanation for this observation remains to be found.

The $\Upsilon(5S)$ also features in the anomalous decay

\be
Bf(\Upsilon(5S) \to B^*B\pi) = (7.3 \pm 2.2 \pm 0.8) \%
\ee
which is ten times higher than expected \cite{bf5s}.

Finally, the decays of $\Upsilon$ to $\Upsilon\pi\pi$ are a rich source of information on nonperturbative hadronisation. Amongst the many curiosities in this area are those presented in Table \ref{UpsTab} \cite{5s}. The difference of two orders of magnitude between $\Upsilon(5S)$ decays and other $\Upsilon$'s is striking. Explanations have invoked final state interactions or exotic nearby resonances \cite{simonov}.

\begin{table}[h]
\caption{$\Upsilon(nS)$ Decays}
\vspace{0.4cm}
\begin{center}
\begin{tabular}{|ll|}
\hline
process &  rate (MeV) \\
\hline
$\Upsilon(5S) \to \Upsilon(1S) \pi\pi$ &  $0.59 \pm 0.04 \pm 0.09$ \\
$\Upsilon(5S) \to \Upsilon(2S) \pi\pi$ &  $0.85 \pm 0.07 \pm 0.16$ \\
$\Upsilon(5S) \to \Upsilon(3S) \pi\pi$ &  $0.52 \pm 0.18 \pm 0.10$ \\
\hline
$\Upsilon(2S) \to \Upsilon(1S) \pi\pi$ &  $0.0060$ \\
$\Upsilon(3S) \to \Upsilon(1S) \pi\pi$ &  $0.0009$ \\
$\Upsilon(4S) \to \Upsilon(1S) \pi\pi$ &  $0.0019$ \\
\hline
\end{tabular}
\end{center}
\label{UpsTab}
\end{table}

\section{Spectroscopy}

Brief observations on recent spectroscopic results follow.

\subsubsection{$Y$'s}

The Belle collaboration has reported the resonances $Y(4350)$ and $Y(4660)$ in $e^+e^- \to \gamma_{\rm ISR} \psi(2S) \pi\pi$ but not in $J/\psi$. It is difficult to understand why the $Y$'s should be visible in $\psi(2S)\pi\pi$ but not in $J/\psi\pi\pi$. However, this phenomenon is reminiscent of the strange effect noted above in $J/\psi$ and $\psi(2S)$ decay to $\pi\pi\pi$. Could $n\pi$ hadronisation somehow reflect radial structure in parent hadrons? 

The $Y(3940)$ was seen by Belle and BaBar in $B \to KY \to K J/\psi \omega$. This state is in a rather crowded mass region, with the $X(3940)$ likely taking a $\chi_{cJ}(2S)$ spot. It is possible it is a threshold effect. However new data from BaBar \cite{bb2} show that the $Y(3940)$ is produced with different strengths in different charge modes:

\be
\frac{Bf(B^0\to K^0 Y)}{Bf(B^+ \to K^+ Y)} = 0.27 \pm 0.25 \pm 0.02.
\ee
This is approximately three standard deviations below what is expected by isospin. The data also calls into question threshold explanations since threshold dynamics and kinematics should be independent of the production mechanism.

\subsubsection{$X(3872)$}

The $X(3872)$ is by now the senior citizen of the charmonium exotics zoo. Although the discovery dates from Belle's paper of 2003 \cite{choi}, it is likely that it was first seen in the PhD research of Tom LeCompte at Fermilab in 1992 \cite{tom}. 

Although Belle's initial observation of the $X$ in the $J/\psi\omega$ decay mode lent strong evidence to the putative molecular nature of this state, this work was never published. It was therefore significant that BaBar has finally observed the $X$ in this decay mode \cite{X-qn}. The collaboration obtained

\be
\frac{Bf(X(3872) \to J/\psi \omega)}{Bf(X(38720 \to J/\psi \pi\pi)} = 0.7 \pm 0.3,
\ee
in agreement with an old model prediction \cite{essX}.

While confirming the $X$ in this channel the collaboration also obtained enough statistics to perform an angular analysis and concluded that the likely quantum numbers of the state are $J^{PC} = 2^{-+}$. If true, this observation kills the molecular interpretation of the $X$ since that hypothesis singles out $1^{++}$ as the only possible quantum numbers. However, as has recently been stressed \cite{kam}, the data are rather sparse and the relative significance of the $1^{++}$ and $2^{-+}$ assignments are quite similar.

The observation of the $X$ in the radiative transitions $X\to \gamma J/\psi$ and $X\to \gamma \psi(2S)$ was important for establishing the charge conjugation of the state and as a diagnostic for its internal structure. Simple arguments in the molecular picture indicate that decays to $\psi(2S)$ (or any other charmonia other than $J/\psi$) should be strongly suppressed \cite{ess2}. Thus the observation of this decay by Babar \cite{babarXg}
is a serious blow to a pure molecular picture of the $X$. A possible resolution is that the $X$ contains substantial $c\bar c$ components \cite{HH}, but recent data from Belle call into question the entire scenario:  Belle report a clear signal of the $X$ in $\gamma J/\psi$ but see no evidence for it in $\gamma \psi(2S)$ \cite{jolanta}.

Finally, it is worth stressing that the relative production of $X$ in charged versus neutral $B$ decays is an important diagnostic for the structure of the $X$. It also serves to test the cusp explanation for the state. If the $X$ signal were due to the kinematics of opening channels it should be seen equally well in charged final states. The fact that charged $X$'s have not been seen is therefore a strong indication that the $X$ is more than a kinematical effect.

\subsubsection{Charged Charmonium States}

The discovery of charged charmonium states by Belle in 2008 unleashed a new wave of interest in exotic charmonia \cite{belleZ}. Such states are of course manifestly exotic in composition and are likely to consist of $c\bar c u \bar d$ (etc) valence quarks. Unfortunately the evidence for the $Z_1(4051)$ and $Z_2(4250)$, seen in $B \to K Z \to K \chi_{c1} \pi$,  is not overwhelming. Alternatively, the signal for the $Z(4430)$ (seen in $B \to KZ \to K \psi(2S) \pi$) is compelling. But, as stressed by the BaBar collaboration, 
identifying resonant structures in the $\psi(2S) \pi$  channel
in three-body decays has to be done
with care because of the possibility that dynamics in
the $K\pi$ channel can create bumps in the $\psi(2S)\pi$  invariant mass distribution. And indeed, BaBar claim that they see no signal for the $Z(4430)$ \cite{noZ}. To paraphrase a noted astronomer, extraordinary claims require extraordinary evidence, and prudence is certainly not to be avoided in this case.

\subsubsection{Charged Bottomonium States}

A few months ago the Belle collaboration reported the observation of narrow charged resonances in
the $\Upsilon(nS)\pi^\pm$ channels with masses of 10610 MeV and 10650 MeV \cite{chargedB}. The favoured quantum numbers are $I^GJ^P = 1^+ 1^+$. These states are just above $B\bar B^*$ (10605 MeV) and $B^*\bar B^*$ (10650 MeV) thresholds respectively. Again, if a resonant character can be verified for these states then exotic multiquark structures must be considered a leading candidate for their composition.

Given the proximity to $B\bar B^*$ and $B^*\bar B^*$ thresholds, it is possible that these states are isovector bound states of these mesons. We assume that a simple pion-exchange model exchange model that is tuned to deuteron properties can provide guidance to $B\bar B^*$ and $B^*\bar B^*$ dynamics. This model was also used by several groups to describe the $X(3872)$.  One finds that the $1^+1^+$ $B^*\bar B^*$ channel is dominated by the $^{2S+1}L_J = {}^5D_1$ state and that it is mildly attractive. Of course the $D$-wave dynamics does not help in forming possible bound states. Thus it is unlikely that the 10650 is a $B^*\bar B^*$ bound state. It may instead be simply a threshold enhancement due to the attractive final state interactions in this channel. Similarly, the $1^+1^+$ $B\bar B^*$ mode experiences a mildly repulsive force in the $^3S_1-{}^3D_1$ channel. Thus it is also unlikely that the 10610 is a bound state (at least within the context of this simple model). Furthermore, the fluke in charmonium that places the open charm $D\bar D^*$ threshold at the $\rho J/\psi$ mass does not occur in the bottomonium spectrum. Note that the 10650 signal is slightly above threshold, in keeping with the weak repulsion found here.

\section{Conclusions}

The interactions and static properties of hadrons provide one of the few experimental routes to exploring aspects of nonperturbative field theory. Unfortunately, there is reason to believe that our tools for this theory can be rudimentary. It is likely that the application of perturbative QCD to exclusive hadronic processes is too naive. Effective field theory approaches show greater promise but are still limited in applicability and sometimes fail in unexpected ways. Lattice gauge theory is progressing rapidly and unquenched results with excited states are beginning to appear. Coupling these states to the continuum remains to be implemented and a theoretical formalism for interpreting  the ensuing results needs to be established. 

Experimental results have been presented that strongly indicate that our understanding of short scale processes and factorisation may be less robust than we would like. The exotic charmonium spectrum continues to grow, but many of the putative states require confirmation, some of which must await a  super B factory. Finally, there is reason to believe that none of the new charged charmonium or bottomonium signals represent new states.  There is plenty to do, and one must look forward to new results from 
JPARC, a super B factory, COMPASS, PANDA/FAIR, BESIII, and the  LHC.

\section*{Acknowledgments}
This research was supported by the U.S. Department of Energy under contract
DE-FG02-00ER41135.

\end{document}